\documentclass[prd,aps,floatfix,nofootinbib, 11 pt]{revtex4}
\usepackage{amsmath}
\usepackage{graphicx,color,epsfig,xcolor}
\usepackage[colorlinks=false]{hyperref}
\usepackage{tensor}

\newcommand{\bea}{\begin{eqnarray}}
\newcommand{\eea}{\end{eqnarray}}
\newcommand{\beq}{\begin{eqnarray}}
\newcommand{\eeq}{\end{eqnarray}}

\begin{document}

\title{Lepton flavor violating decays of $\mathbf{\mu}$ and $\mathbf{\tau}$
leptons
in a gauge group $\mathbf{\ SU(2)_L\times SU(2)_R\times SU(2)_Y} $ }
\author{Fayyazuddin}
\affiliation{National Centre for Physics,\\
Quaid-i-Azam University Campus, Islamabad 45320, Pakistan}
\date{\today }

\begin{abstract}
The electroweak unification group $ SU(2)_L\times SU(2)_R\times SU(2)_Y $ is proposed for the charged lepton flavor violating decays of the muon ($\mu$) and tau ($\tau$) leptons.  The group $SU(2)_Y$ is in the lepton space. The left-handed leptons and anti-leptons are assigned to the fundamental representation $(2,2,\bar{2})$ of the semi-simple group. The gauge group $SU(2)_Y$ is spontaneously broken to $U(1)_{Y_1}$, where $Y_1=-L=\pm1$ is the hypercharge, by introducing a scalar multiplet $\Sigma$ which belongs to the triplet representation 3 of the $SU(2)_Y$ and is singlet under $SU(2)_L\times SU(2)_R$. At this stage charged vector bosons $Y^\pm$ of $SU(2)_Y$ which mediate the lepton flavor violating decays acquire masses and are decoupled with one Higgs scalar $H_\Sigma^0$. The residual group $SU(2)_L\times SU(2)_R\times U(1)_{Y_1}$ has all the features of the left-right electroweak unification group extensively studied in the literature. The probability for lepton flavor violating decays is $\left(\frac{\sin^2\theta_W}{1-2\sin^2\theta_W}\right)^2\left(\frac{m_{W_L}}{m_Y}\right)^4$.
\end{abstract}

\maketitle




\section{Introduction}

In the standard model, lepton number and baryon number are conserved, i.e. $\Delta
L=0$  and $\Delta B=0$. {Bounds on the lifetimes of electron and proton are} 
\begin{equation}
\tau _{e}>4.6\times 10^{26}\text{years}, \hspace{1cm}\tau _{p}>10^{31}\text{years}.
\label{1}
\end{equation}%
For the leptons, there is another conservation law, viz the lepton number for
each generation is conserved. No process with $\Delta L_{e}\neq 0$, $\Delta
L_{\mu }\neq 0$ and   $\Delta L_{\tau}\neq 0$ is allowed. The left-handed current for weak decays in the
standard model is 
\begin{align}
J^{+\mu }_{\text{lepton}} &=\big(\bar{\nu}_{e},\bar{\nu}_{\mu },\bar{\nu}_{\tau
}\big)\gamma ^{\mu }(1-\gamma ^{5})\left( 
\begin{array}{c}
e \\ 
\mu \\ 
\tau%
\end{array}%
\right) 
=\Big[\bar{\nu}_{e}\gamma ^{\mu }(1-\gamma ^{5})e+\cdots \Big]  \label{2}\\
J^{+\mu }_{\text{quark}} &=\big(\bar{u},\bar{c},\bar{t}\big)\gamma ^{\mu }(1-\gamma
^{5})V_{CKM}\left( 
\begin{array}{c}
d \\ 
s \\ 
b%
\end{array}%
\right) =\Big[\bar{u}\gamma ^{\mu }(1-\gamma ^{5})(V_{ud}d+V_{us}s+V_{ub}b)+\cdots \Big]
\label{3}
\end{align}%
where $V_{CKM}$ is the Cabibbo--Kobayashi--Maskawa (CKM) matrix. Unlike the quark-sector, where the flavor changing weak decays are allowed, in
the lepton-sector no lepton flavor changing decays are allowed.

In the standard model, the neutrinos are {only} left-handed and {hence} are massless. For
massless neutrinos, no linkage between three generations. Mixing between
neutrinos is  possible if all the neutrinos are not massless. Mixing requires
that mass eigenstates $\nu _{i}$, $(i=1,2,3)$ are different from flavor
eigenstates. In this case the oscillations are possible. In the neutrino
oscillations, $\nu _{e}\rightarrow \nu _{\mu }$ and $\nu _{\mu }\rightarrow \nu
_{\tau }$ have been experimentally observed \cite{1, 2} .

One has to go beyond the standard model to explore the {charged} lepton flavor
violating decays. In this context, the electroweak unification gauge group $%
SU(2)_{L}\times SU(2)_{R}\times SU(2)_Y$ is proposed, the group $SU(2)_Y$
is in the leptonic space. The concept of isospin in the leptonic space was
first introduced in 1975 \cite{3}. The left-handed leptons and antileptons are
assigned to the fundamental representation $(2,2,\bar{2})$ of the gauge
group: 
\begin{equation}
\Psi =
{
\tensor[_{SU(2)_L}]{\Bigg\downarrow}{}
\overset{\xleftarrow{~~SU(2)_Y~~}}{\begin{pmatrix}
\nu _{n} & e_{m}^{c} \\ 
e_{n} & -N_{m}^{c}%
\end{pmatrix}} \Bigg\downarrow_{ SU(2)_R}
}
\label{4}
\end{equation}
where $n$ and $m$ are the  flavor indices and the superscript $c$ denotes the charge-conjugation. The multiplets $\left(\nu_n, ~ e_n\right)^{\text{T}}_L$ and  $\left( 
e^c_m, ~-N^c_m\right)^{\text{T}}_L$ belong to the fundamental representations of $S U(2)_L$ and $S
U(2)_R$, respectively, whereas two doublets $(\nu_n , e^c_m)_L$ and $(e_n ,
-N^c _m)_L$ belong to the representation $\bar{2}$ of $S U(2)_Y$. There are
three sets of vector bosons $(W^{\pm}_L , W^0_{L \mu})$, $(W^{\pm}_{R \mu} ,
W^0_{R \mu})$ and $(Y^{\pm}_\mu , Y^0_{\mu})$ belonging to the adjoint
representation of each $S U(2)$ gauge group.
Out of six charged vector bosons, the four $W^{\pm}_{L \mu}$ , $W^{\pm}_{R
\mu}$ are coupled to the left-handed and right-handed weak currents $J^{\pm
\mu}_L$ and $J^{\pm \mu}_R$ respectively. The remaining two charged vector
bosons $Y^{\pm} _\mu$ are coupled to the lepton flavor violating currents $%
J^{\pm \mu}_Y$. The gauge bosons $Y^{\pm}_\mu$ mediate the lepton number $L_e$, $L_\mu$ and $%
L_\tau$ violating processes. The linear combinations of three
neutral vector bosons $W^0_{L \mu}$, $W^0_{R \mu}$ and $Y^0_{\mu}$ give three
physical vector bosons $A_\mu$, $Z_\mu$ and $Z^{\prime}_\mu$ coupled is the
electromagnetic current $J^{e m \mu}$, the weak neutral currents $J^{Z
\mu}$ and the $J^{Z^\prime \mu}$, respectively.

{Note that for the gauge group $SU(2)$, the representations $2$ and $\bar{2}$ are equivalent and it is
anomaly free unlike $SU(N)$ for $N>2$ gauge groups which are not anomaly free. Hence,} the gauge group $SU(2)_{L}\times
SU(2)_{R}\times SU(2)_Y$ is anomaly free.

\section{Interaction Lagrangian}

The gauge invariant Lagrangian for the fundamental representation $(2,2,\bar{%
2})$ is given by 
\begin{equation}
{\cal L}=\text{Tr}[i\bar{\Psi}\gamma ^{\mu }\nabla _{\mu }\Psi ], 	 \label{5}
\end{equation}%
with $\Psi$ defined in Eq. \eqref{4} and
\begin{equation}
\nabla _{\mu }=\partial _{\mu }+\frac{i}{2}\tau \cdot W_{L\mu }+\frac{i}{2}\tau\cdot W_{R\mu }-\frac{i}{2}(\tau \cdot Y_{\mu })^{\dagger }, \label{6}
\end{equation}%
where $\tau^a$ are the Pauli matrices. Thus the interaction Lagrangian 
\begin{align}
{\cal L}_{\text{int}} &=\frac{i}{2}\Bigg[(\bar{\nu _{n}},\bar{e_{n}})_{L}\gamma ^{\mu
}\left( 
\begin{matrix}
W^0_{L_{\mu }} & \sqrt{2}W_{L_{\mu }}^{+} \\ 
\sqrt{2}W_{L_{\mu }}^{-} & -W^0_{L_{\mu }}%
\end{matrix}%
\right) \left( 
\begin{matrix}
\nu _{n} \\ 
e_{n}%
\end{matrix}%
\right) _{L}+(\bar{e}_{n}^{c},-\bar{N}_{n}^{c})_{L}\gamma ^{\mu }\left( 
\begin{matrix}
W^0_{R_{\mu }} & \sqrt{2}W_{R_{\mu }}^{+} \\ 
\sqrt{2}W_{R_{\mu }}^{-} & -W^0_{R_{\mu }}%
\end{matrix}%
\right) \left( 
\begin{matrix}
e_{n}^{c} \\ 
-N_{n}^{c}%
\end{matrix}%
\right) _{L}  \notag \\
&\hspace{-0.1cm}-(\bar{\nu _{n}},\bar{e_{m}^{c}})_{L}\gamma ^{\mu }\left( 
\begin{matrix}
Y^0_{\mu } & \sqrt{2}Y_{\mu }^{-} \\ 
\sqrt{2}Y_{\mu }^{+} & -Y^0_{\mu }%
\end{matrix}%
\right) \left( 
\begin{matrix}
\nu _{n} \\ 
e_{m}^{c}%
\end{matrix}%
\right) _{L}-(\bar{e}_{n},-\bar{N}_{m}^{c})_{L}\gamma ^{\mu }\left( 
\begin{matrix}
Y^0_{\mu } & \sqrt{2}Y_{\mu }^{-} \\ 
\sqrt{2}Y_{\mu }^{+} & -Y^0_{\mu }%
\end{matrix}%
\right) \left( 
\begin{matrix}
e_{n} \\ 
-N_{m}^{c}%
\end{matrix}%
\right) _{L}\Bigg]  \label{intlag}
\end{align}%
From the above equation we can separate the charged and neutral parts of the interaction Lagrangian as
\begin{align}
{\cal L}_{\text{int}}^{\text{charge}} =&-\frac{1}{2\sqrt{2}}\Big\{g\big[\bar{\nu _{n}}%
\gamma ^{\mu }(1-\gamma ^{5})e_{n}W_{L_{\mu }}^{+}+\bar{N_{n}}\gamma ^{\mu
}(1+\gamma ^{5})e_{n}W_{R_{\mu }}^{+}+\text{h.c.}\big]  \notag \\
&-g_{Y}\big[(\bar{e}_{m}^{c}\gamma ^{\mu }(1-\gamma ^{5})\nu _{n}-\bar{N}%
_{m}^{c}\gamma ^{\mu }(1-\gamma ^{5})e_{n})Y_{\mu }^{+}+\text{h.c.}\big]\Big\}
\label{charged-lag}\\
{\cal L}_{\text{int}}^{\text{neutral}} =&-\frac{1}{4}\Big\{g\big[(\bar{\nu _{n}}\gamma
^{\mu }(1-\gamma ^{5})\nu _{n}-\bar{e_{n}}\gamma ^{\mu }(1-\gamma
^{5})e_{n})W^0_{L_{\mu }}+(\bar{N_{n}}\gamma ^{\mu }(1+\gamma ^{5})N_{n} -\bar{e_{n}}\gamma ^{\mu }(1+\gamma ^{5})e_{n})W^0_{R_{\mu }}\big]
\notag \\
&-g_{Y}\big[\bar{%
\nu _{n}}\gamma ^{\mu }(1-\gamma ^{5})\nu _{n}+\bar{e_{n}}\gamma ^{\mu
}(1+\gamma ^{5})e_{n}  +\bar{e_{n}}\gamma ^{\mu }(1-\gamma ^{5})e_{n}+\bar{N_{n}}\gamma ^{\mu
}(1_{+}\gamma ^{5})N_{n}\big]Y^0_{\mu }\Big\}  \label{neutral-lag}
\end{align}%
In order to express the ${\cal L}_{\text{int}}^{\text{neutral}}$ in terms of physical vector bosons 
$A_{\mu }$, $Z_{\mu }$ and $Z_{\mu }^{\prime }$, we note that electric charge $%
Q $ is given by 
\begin{align}
Q&=I_{3L}+I_{3R}+I_{3Y} \\
&~~~~~g~~~~~~~g~~~~~~~g_{Y}\notag \\
&~~~~W^0_{L_{\mu }}~~~W^0_{R_{\mu }}~~~Y^0_{\mu } \notag
\end{align}%
{Below we define the gauge bosons and the couplings in the mass eigenbases,}
\begin{align}
\frac{A_{\mu }}{e} &=\frac{W^0_{L_{\mu }}}{g}+\frac{B_{\mu }}{g^{\prime }} ,
&\frac{B_{\mu }}{g^{\prime }} &=\frac{W^0_{R_{\mu }}}{g}+\frac{Y^0_{\mu }}{%
g_{Y}}, \hspace{2cm} \frac{Z_{\mu }^{\prime }}{g^{\prime }} =\frac{W^0_{R_{\mu }}}{g_{Y}}-\frac{%
Y_{0\mu }}{g} , \notag \\
\frac{1}{e^{2}} &=\frac{1}{g^{2}}+\frac{1}{g^{\prime 2}}, &\frac{1}{g^{\prime 2}} &=\frac{1}{g^{2}}+\frac{1}{g_{Y}^{2}},  \notag \\
\frac{e}{g} &=\sin \theta _{W}, &\frac{e}{g^{\prime }}& =\cos \theta _{W},\hspace{2.8cm}g_{Y}=\frac{g\tan \theta _{W}}{%
\sqrt{1-\tan ^{2}\theta _{W}}}.  \label{wbza}
\end{align}%
{From the above definitions,} one can obtain $W^0_{L_{\mu }}$, $W^0_{R_{\mu }}$
and $Y^0_{\mu }$ in terms of the physical vector bosons $A_{\mu }$, $Z_{\mu }$
and $Z_{\mu }^{\prime }$ as: 
\begin{align}
gW^0_{L_{\mu }} &=eA_{\mu }+g\cos {\theta }_{W}Z_{\mu }  \notag \\
gW^0_{L_{\mu }}-g_{Y}Y_{0\mu } &=\frac{g}{\cos \theta _{W}}Z_{\mu }+g\frac{%
\tan ^{2}\theta _{W}}{\sqrt{1-\tan ^{2}\theta _{W}}}Z_{\mu }^{\prime } 
\notag \\
gW^0_{R_{\mu }} &=eA_{\mu }-g\frac{\sin ^{2}\theta _{W}}{\cos \theta _{W}}%
Z_{\mu }+g\sqrt{1-\tan ^{2}\theta _{W}}Z_{\mu }^{\prime }  \notag \\
g_{Y}Y^0_{\mu }z &=eA_{\mu }-g\frac{\sin ^{2}\theta _{W}}{\cos \theta _{W}}%
Z_{\mu }-g\frac{\tan ^{2}\theta _{W}}{\sqrt{1-\tan ^{2}\theta _{W}}}Z_{\mu
}^{\prime }  \notag \\
gW^0_{R_{\mu }}-g_{Y}Y^0_{\mu } &=\frac{g}{\sqrt{1-\tan ^{2}\theta _{W}}}%
Z_{\mu }^{\prime } \notag \\
g(W^0_{L\mu }-W^0_{R\mu }) &=\frac{g}{\cos \theta _{W}}Z_{\mu }-g\sqrt{1-\tan
^{2}\theta _{W}}Z_{\mu }^{\prime }   \label{wzrelations}
\end{align}%
Using above relations, the neutral current interaction Lagrangian is given
by 
\begin{align}
{\cal L}_{\text{int}}^{\text{neutral}} &=\frac{-1}{4}\bigg\{e\Big[-4\bar{e_{n}}\gamma ^{\mu
}e_{n}\Big]A_{\mu }+g\Big[\bar{\nu _{n}}\gamma ^{\mu }(1-\gamma ^{5})-\bar{e_{n}}%
\gamma ^{\mu }(1-\gamma ^{5})e_{n}  \notag \\
&-4\sin ^{2}\theta _{W}(-\bar{e_{n}\gamma ^{\mu }e_{n}})\Big]\frac{Z_{\mu }}{%
\cos \theta _{W}}+g\Big[(\bar{N_{n}}\gamma ^{\mu }(1+\gamma ^{5})N_{n}-\bar{e_{n}%
}\gamma ^{\mu }(1+\gamma ^{5})e_{n})  \notag \\
&+\tan ^{2}\theta _{W}(\bar{\nu _{n}}\gamma ^{\mu }(1-\gamma ^{5})\nu _{n}-%
\bar{e_{n}}\gamma ^{\mu }(1-\gamma ^{5})e_{n}+4\bar{e_{n}}\gamma ^{\mu
}e_{n})\Big]\frac{Z_{\mu }^{\prime }}{\sqrt{1-\tan ^{2}\theta _{W}}}\bigg\}
\label{neutral-lag1}
\end{align}%
We conclude from Eq. (\ref{charged-lag}) and Eq. (\ref{neutral-lag1}), that
except for lepton number violating term coupled to $Y_{\mu }^{\pm }$, we get
exactly the same result for the lepton sector as those given by the
left-right symmetric gauge group $SU(2)_{L}\times SU(2)_{R}\times
U(1)_{Y_{1}}$, see for instance \cite{4}.

\section{Spontaneous breaking of the gauge group $\mathbf{SU(2)_L\times
SU(2)_R\times SU(2)_Y}$}

{In the first stage, the group $SU(2)_Y$ is spontaneously broken to $%
U(1)_{Y_{1}}$, where $Y_{1}$ is the hypercharge, by a scalar multiplet $%
\Sigma $ which belong to singlet representation of $SU(2)_{L}$, $SU(2)_{R}$
and to triplet representation of $SU(2)_Y$, i.e. $\Sigma= (1,1,3)$ and can be written in the following form
\begin{equation}
\Sigma =\left( 
\begin{array}{c}
H_{\Sigma }^{+} \\ 
 v_{\Sigma}+H_{\Sigma }^{0}  \\ 
H_{\Sigma }^{-}%
\end{array}%
\right),   \label{3.2}
\end{equation}%
where $\langle\Sigma\rangle \equiv (0, v_\Sigma, 0)^{\text{T}}$ is the vacuum expectation value (vev) of $\vec\Sigma$.}
The mass term is given by 
\begin{align}
{\cal L}^{Y^\pm}_{\text{mass}} &=-\frac{1}{4}g_{Y}^{2}\left[ \big(\langle \vec{\Sigma}\rangle \cdot\langle \vec{\Sigma}\rangle \big)\big(\vec{Y}^{\mu }\cdot\vec{Y}_{\mu }\big)
-\big(\langle \vec{\Sigma}\rangle \cdot\vec{Y}^{\mu }\big)\big(\langle 
\vec{\Sigma}\rangle \cdot\vec{Y}_{\mu }\big)\right]   \notag \\
&=-\frac{1}{4}g_{Y}^{2}V^{2}\left[ 2Y^{+\mu }Y_{\mu }^{-}\right] ,
\label{3.4}
\end{align}
where the mass of the gauge bosons $Y^\pm$ is given by
\begin{equation}
m_{Y\pm }^{2}=\frac{1}{4}g_{Y}^{2}(2v_\Sigma^{2})=\frac{1}{4}g^{2}\frac{\tan
^{2}\theta _{W}}{1-\tan ^{2}\theta _{W}}\left( 2v_\Sigma^{2}\right). \label{mYmass}
\end{equation}%
The would be Goldstone bosons $H_{\Sigma }^{\pm }$ have been absorbed in $%
Y_{\mu }^{\pm }$ to give them longitudinal components and masses. The vector
bosons $Y_{\mu }^{\pm }$ are decoupled with one heavy Higgs scalar $%
H_{\Sigma }^{0}$ and the electroweak unification group $SU(2)_{L}\times
SU(2)_{R}\times SU(2)_Y$ is broken to $SU(2)_{L}\times SU(2)_{R}\times
U(1)_{Y_{1}}$. We are left with seven massless vector bosons $\left( W_{L\mu }^{\pm
},W_{L\mu }^{0}\right) $, $\left( W_{R\mu }^{\pm },W_{R\mu }^{0}\right) $
and a singlet $Y_{\mu }^{0}$ belonging to $SU(2)_{L}$, $SU(2)_{R}$ and $%
U(1)_{Y_{1}}$, respectively and the two doublets 
\begin{equation}
\left( 
\begin{array}{c}
\nu _{n} \\ 
e_{n}%
\end{array}%
\right) _{L}~,~~~~\left( 
\begin{array}{c}
e_{n}^{c} \\ 
-N_{n}^{c}%
\end{array}%
\right) _{L}  \label{3.5}
\end{equation}%
belonging to representation $2$ of $SU(2)_{L}$ and $SU(2)_{R}$ with hypercharge 
$Y_{1}=-L=\mp 1$. The singlet vector boson $Y_{\mu }^{0}=B_{1\mu }$, $%
g_{Y}=g_{1}$.

In the second stage, the gauge group $SU(2)_{L}\times SU(2)_{R}\times U(1)_{Y_{1}}$ (left-right symmetric group) is
spontaneously broken to $U(1)_{em }$ by three sets of scalars \cite{4}: 
\begin{align}
\Delta _{R}:(1,2,2)&=\left( 
\begin{matrix}
\eta ^{+} & \eta ^{++} \\ 
\eta ^{\sigma } & -\eta ^{+}%
\end{matrix}%
\right) , &\left\langle \Delta _{R}\right\rangle &=\left( 
\begin{matrix}
0 & 0 \\ 
v _{R}/2 & 0%
\end{matrix}%
\right)  \label{3.6}\\
\Delta _{L}:(2,1,2)&=\left( 
\begin{matrix}
\chi ^{+} & \chi ^{++} \\ 
\chi ^{0} & -\chi ^{+}%
\end{matrix}%
\right) , &\left\langle \Delta _{L}\right\rangle &=\left( 
\begin{matrix}
0 & 0 \\ 
v _{L}^{\prime }/\sqrt{2} & 0%
\end{matrix}%
\right) \approx 0  \label{3.7}\\
\phi :(2,2,0)&=\left( 
\begin{matrix}
\phi ^{0} & \phi ^{+} \\ 
\phi ^{-} & -\phi ^{0}%
\end{matrix}%
\right) , &\left\langle \phi \right\rangle &=\left( 
\begin{matrix}
\kappa & 0 \\ 
0 & \kappa ^{\prime }%
\end{matrix}%
\right)  \label{3.8}
\end{align}%
The multiplet $\Delta _{R}$, generates the mass terms for the $W_R^\pm$ and $Z^\prime$ gauge bosons by using Eq (\ref{wzrelations}) as 
\begin{align}
{\cal L}^{W^\pm_R,Z^\prime}_{\text{mass}}&=-\frac{1}{8}v_{R}^{2}\left[ 2g^{2}W_{R}^{+\mu }W_{R\mu }^{-}+2\left(
gW_{R}^{0\mu }-gB_{1}^{\mu }\right) \left( gW_{R\mu }^{0}-gB_{1\mu }\right) %
\right] \notag\\
&=-\frac{1}{8}g^{2}v_{R}^{2}\left[ 2W_{R}^{+\mu }W_{R\mu }^{-}+\frac{2}{%
1-\tan ^{2}\theta _{W}}Z^{\prime \mu }Z_{\mu }^{\prime }\right] . \label{3.9}
\end{align}%
Hence 
\begin{equation}
m_{W_{R}^{\pm }}^{2}=\frac{1}{4}g^{2}v_{R}^{2}, \hspace{1cm}m_{Z^{\prime }}^{2}=%
\frac{1}{4}\frac{2g^{2}v_{R}^{2}}{1-\tan ^{2}\theta _{W}}  \label{3.10}
\end{equation}%
At this stage the left-right symmetric group is broken to $%
SU(2)_{L}\times U(1)_{Y}$. The scalar multiplet $\phi $ breaking this group
to $U(1)_{em}$. {The multiplet $\phi $ generate the mass term for the $W^\pm_L$ and $Z$ as,
\begin{align}
{\cal L}^{W^\pm_L,Z}_{\text{mass}}&=\frac{1}{4}g^{2}
\left( \kappa ^{2}+\kappa ^{\prime 2}\right) \left[ \left( 2W_{L}^{+\mu
}W_{L\mu }^{-}+2W_{R}^{+\mu }W_{R\mu }^{-}\right) +\left( W_{L}^{0\mu
}-W_{R}^{0\mu }\right) \left( W_{L\mu }^{0}-W_{R\mu }^{0}\right) \right] \notag\\ 
&\hspace{.5cm}-g^2\kappa \kappa ^{\prime }\left( W_{L}^{+\mu }W_{R\mu }^{-}+W_{R}^{+\mu
}W_{L\mu }^{-}\right)
  \label{3.12}\\
&=\frac{1}{4}g^{2}\kappa ^{2}\bigg[ 2W_{L}^{+\mu }W_{L\mu }^{-}+2W_{R}^{+\mu
}W_{R\mu }^{-}+\frac{1}{\cos ^{2}\theta _{W}}Z^{\mu }Z_{\mu }-\frac{\sqrt{%
1-\tan ^{2}\theta _{W}}}{\cos \theta _{W}}\left( Z^{\prime \mu }Z_{\mu
}+Z^{\mu }Z_{\mu }^{\prime }\right) \notag\\
& \hspace{1.5cm}+\left( 1-\tan ^{2}\theta _{W}\right)
Z^{\prime \mu }Z^{\prime}_{\mu }\bigg]
\end{align}
where in the last step we used  Eq. (\ref{wzrelations})  and the fact that  $\kappa ^{\prime}\ll\kappa $ (which one can select). }
Hence with $\kappa ^{\prime }\ll\kappa \ll v _{R}$, and $\kappa =v _{L}/
\sqrt{2}$, we get
\begin{equation}
m_{W_{L}^{\pm }}=\frac{1}{4}g^{2}v _{L}^{2}, \hspace{1cm}m_{Z}^{2}=\frac{1}{4}\frac{%
g^{2}}{cos^{2}\theta _{W }}v _{L}^{2}  \label{3.13}
\end{equation}%

The scalar multiplet $\Delta _{R}$ gives the Majorana mass term to the
right-hand neutrino $N_{n}$: 
\begin{align}
{\cal L}^{\text{Majorana}}_{\text{mass}}&=-\left( e_{n}^{T},-N_{n}^{c^{T}}\right) _{L}C^{-1}i\tau _{2}\langle
\Delta _{R}\rangle^{\dagger} \left( 
\begin{array}{c}
e_{n} \\ 
-N_{n}^{c}%
\end{array}%
\right) _{L}+\text{h.c.}  \notag \\
&=-\left( e_{n}^{T},-N_{n}^{c^{T}}\right) _{L}C^{-1}\left( 
\begin{matrix}
0 & 0 \\ 
0 & v _{R}/2%
\end{matrix}%
\right) \left( 
\begin{array}{c}
e_{n} \\ 
-N_{n}^{c}%
\end{array}%
\right) _{L}+\text{h.c.}  \notag \\
&=-\frac{v _{R}}{2}\left[ N_{nL}^{c^{T}}C^{-1}N_{nL}^{c}-\bar{N}_{nL}^{c}C%
\bar{N}_{nL}^{c^{T}}\right]  \notag \\
&=-\frac{v _{R}}{2}\Big[ N_{nR}^{T}C^{-1}N_{nR}+\text{h.c.}\Big].  \label{3.14}
\end{align}%
{The multiplet $\phi $ generates Dirac masses for the leptons. The Dirac mass term for the leptons is 
\begin{align}
{\cal L}^{\text{Dirac}}_{\text{mass}}&=-\left[ h_{1l_{n}}\left( \nu _{n}^{T},e_{n}^{+}\right) C^{-1}\left\langle
\phi \right\rangle i\tau _{2}\left( 
\begin{array}{c}
e_{n}^{c} \\ 
-N_{n}^{c} \\ 
\end{array}%
\right) _{L}+h_{2l_{n}}\left( \nu _{n}^{T},e_{n}^{+}\right) C^{-1}i\tau
_{2}\left\langle \phi \right\rangle \left( 
\begin{array}{c}
e_{n}^{c} \\ 
-N_{n}^{c} \\ 
\end{array}%
\right) _{L}\right] +h.c.  \notag \\
&=-\Big[-h_{1l_{n}}\left( \kappa \nu _{nL}^{T}C^{-1}N_{nL}^{c}+h_{1}\kappa
^{\prime }e_{nL}^{T}C^{-1}e_{nL}^{c}\right) -h_{2l_{n}}\left( \kappa ^{\prime }\nu _{nL}^{T}C^{-1}N_{nL}^{c}+\kappa
e_{nL}^{T}C^{-1}e_{nL}^{c}\right) \Big]+\text{h.c.}  \notag \\
&=-\Big[ \left( h_{1l_{n}}\kappa +h_{2l_{n}}\kappa ^{\prime }\right) \bar{N}%
_{nR}\nu _{nL}+\left( h_{1l_{n}}\kappa ^{\prime }+h_{2l_{n}}\kappa \right)
\left( \bar{e}_{nR}e_{nL}\right) \Big] +\text{h.c.}  \notag\\
&=-\Big[ h_{1l_{n}}\kappa \left( \bar{N}_{nR}\nu _{nL}+\text{h.c.}\right)
+h_{2l_{n}}\kappa \left( \bar{e}_{nR}e_{nL}+\text{h.c.}\right) \Big]  \notag \\
&=-\frac{v _{L}}{\sqrt{2}}\Big[ h_{1l_{n}}\left( \bar{\nu}%
_{nL}N_{nR}+\text{h.c.}\right) +h_{2l_{n}}\left( \bar{e}_{nL}e_{nR}+\text{h.c.}\right) %
\Big], \label{3.16}
\end{align}%
above in the second-last line above we used the approximation $\kappa ^{\prime }\ll \kappa \ll v _{R}$.}
On diagonalization, it gives Majorana mass terms $m_{\nu _{Ln}}=\frac{%
m_{l n}^{2}}{4M_{N_{Rn}}}$. Moreover, the multiplet $\phi$ also generate the quark
masses, the mass term for quarks: 
\begin{equation}
{\cal L}^{\text{quark}}_{\text{mass}}=-\frac{v _{L}}{\sqrt{2}}\left[ h_{1q_n}\left( \bar{u}_{nL}u_{nR}+h.c.\right)
+h_{2q_n}\left( \bar{d}_{nL}d_{nR}+h.c.\right) \right]  \label{3.17}
\end{equation}%

We end this section with the following remark. The content of the gauge vector
bosons and breaking of the gauge group $SU(2)_{L}\times SU(2)_{R}\times
U(1)_{Y_{1}}$ by the Higgs scalar multiplets are characteristic of the gauge
groups and are independent of the fermionic content of the model. In the
left-right symmetric gauge group $SU(2)_{L}\times SU(2)_{R}\times
U(1)_{Y_{1}}$, with $Y_{1}=B-L$ considered in Ref.~\cite{4} the left-handed
quarks and antiquarks doublets 
\begin{equation*}
\left( 
\begin{array}{c}
u_{n} \\ 
d_{n}^{\prime }%
\end{array}%
\right) _{L},\hspace{1cm}\left( 
\begin{array}{c}
d_{n}^{\prime c} \\ 
-u_{n}^{c}%
\end{array}%
\right) _{L}
\end{equation*}%
have $Y_{1}=B$, $Y_{1}=\pm \frac{1}{3}$ whereas lepton (antilepton)
multiplets have $Y_{1}=-L$, $Y_{1}=\mp 1$ as given in Eq. (\ref{3.5}). The
vector bosons $W_{L\mu }^{\pm }$ and $W_{R\mu }^{\pm }$ are coupled to
leptons and quarks with no difference, where as the vector boson $B_{1\mu }$
associated with $U(1)_{Y_{1}}$ is coupled to leptons with $Y_{1}=-L$ and to
quarks with $Y_{1}=B$. The coupling of Higgs scalar with $Y_{1}=0$  and $Y_1 = 2$ are 
coupled to all the fermions of the group.

\section{Effective Lagrangian for charged lepton number violating decays}

In the Standard Model (SM) charged lepton ($\mu $ and $\tau $) decays are mediated by the
vector bosons $W_{L\mu }^{\pm }$. From the first term of Eq. (\ref%
{charged-lag}), the effective Lagrangian for these decays is given by 
\begin{equation}
{\cal L}^{\text{SM}}_{\text{eff}}=\frac{G_{F}}{\sqrt{2}}\big[\bar{\nu}_{n}\gamma ^{\mu }(1-\gamma
^{5})e_{n}\big]\big[ \bar{e}_{m}\gamma _{\mu }(1-\gamma _{5})\nu _{m}\big],
\label{eff1}
\end{equation}%
where $G_{F}=\sqrt{2}g^{2}/8m_{W}^{2}$, is the Fermi constant. After Fierez reordering 
\begin{equation}
{\cal L}^{\text{SM}}_{\text{eff}}=\frac{G_{F}}{\sqrt{2}}\big[\bar{e}_{m}\gamma ^{\mu }(1-\gamma ^{5})e_{n}\big]%
\big[ \bar{\nu}_{n}\gamma _{\mu }(1-\gamma _{5})\nu _{m}\big].
\label{eff-lag-1}
\end{equation}%
In particular, for $\mu ^{-}\rightarrow \nu_{\mu }+e^{-}+\bar{\nu} _{e}$: 
\begin{equation}
{\cal L}^{\text{SM}}_{\text{eff}}(\mu\text{-decay})=\frac{G_{F}}{\sqrt{2}}\big[\bar{e}\gamma ^{\mu }(1-\gamma ^{5})\mu \big]\big[
\bar{\nu _{\mu }}\gamma _{\mu }(1-\gamma _{5})\nu _{e}\big].
\label{eff-lag1}
\end{equation}%
The effective Lagrangian for the charged lepton flavor violating (LFV) decays mediated by $Y_{\mu
}^{\pm }$:%
\begin{equation}
{\cal L}^{\text{LFV}}_{\text{eff}}=\frac{G_{Y}}{\sqrt{2}}\big[\bar{e}_{m}^{c}\gamma ^{\mu }(1-\gamma
^{5})\nu _{n}\big]\big[ \bar{\nu}_{n^{\prime }}\gamma _{\mu }(1-\gamma
_{5})e_{m^{\prime }}^{c}\big], \label{eff2}
\end{equation}%
with 
\begin{equation*}
\frac{G_{Y}}{\sqrt{2}}=\frac{g_{Y}^{2}}{8m_{Y}^{2}}=\frac{g^{2}}{8}\frac{%
\tan ^{2}\theta _{W}}{1-\tan ^{2}\theta _{W}}\frac{1}{m_{Y}^{2}}.
\end{equation*}%
Fierez reordering gives%
\begin{align}
{\cal L}^{\text{LFV}}_{\text{eff}} &=\frac{G_{Y}}{\sqrt{2}}\big[\bar{e}_{m}^{c}\gamma ^{\mu }(1-\gamma
^{5})e_{m^{\prime }}^{c}\big]\big[ \bar{\nu}_{n^{\prime }}\gamma _{\mu
}(1-\gamma _{5})\nu _{n}\big]  \notag \\
&=-\frac{G_{Y}}{\sqrt{2}}\big[e_{m}^{T}C^{-1}\gamma ^{\mu }(1-\gamma ^{5})C\bar{e}%
_{m^{\prime }}^{T}\big]\big[ \bar{\nu}_{n^{\prime }}\gamma _{\mu }(1-\gamma
_{5})\nu _{n}\big]  \notag \\
&=-\frac{G_{Y}}{\sqrt{2}}\big[\bar{e}_{m^{\prime }}\gamma ^{\mu }(1+\gamma
^{5})e_{m}\big]\big[ \bar{\nu}_{n^{\prime }}\gamma _{\mu }(1-\gamma _{5})\nu
_{n}\big].  \label{eff3}
\end{align}%
There is a choice of the possible assignments of three generations of
leptons. The natural assignment is as follow:
\begin{equation}
(i)\hspace{1cm}D(e):D(123)=\left( 
\begin{matrix}
\nu _{e} & e^{+} \\ 
e^{-} & -N_{e}^{c}%
\end{matrix}%
\right) _{L}, ~\left( 
\begin{matrix}
\nu _{\mu } & \mu ^{+} \\ 
\mu ^{-} & -N_{\mu }^{c}%
\end{matrix}%
\right) _{L}, ~\left( 
\begin{matrix}
\nu _{\tau } & \tau ^{+} \\ 
\tau ^{-} & -N_{\tau }^{c}%
\end{matrix}%
\right) _{L}
\end{equation}%
For the assignment $(i)$, the $\mu $ and $\tau $ decays are as follows%
\begin{align}
\mu ^{-} &\rightarrow \bar{\nu}_{\mu }+e^{-}+\nu _{e}, &\Delta L_{\mu
}&=-2,\hspace{1cm}\Delta L_{e}=2,\hspace{1cm}\nu _{e}\leftrightarrow \nu _{\mu }, \\
\tau ^{-} &\rightarrow \bar{\nu}_{\tau }+\mu ^{-}+\nu _{\mu },&\Delta
L_{\tau }&=-2,\hspace{1cm}\Delta L_{\mu }=2,\hspace{1cm}\nu _{\mu }\leftrightarrow \nu
_{\tau }, \\
\tau ^{-} &\rightarrow \bar{\nu}_{\tau }+e^{-}+\nu _{e}, &\Delta L_{\tau
}&=-2,\hspace{1cm}\Delta L_{e}=2,\hspace{1cm}\nu _{e}\leftrightarrow \nu _{\tau }.
\end{align}%
These decays stimulate the neutrino oscillations $\nu _{e}\rightarrow \nu
_{\mu }$ in the $\mu $-decay and $\nu _{\mu }\rightarrow \nu _{\tau }$,~$\nu
_{\tau }\rightarrow \nu _{e}$ in the $\tau $-decay.

The lepton flavor violating effective Lagrangian for $\mu ^{-}\rightarrow \bar{\nu}_{\mu
}+e^{-}+\nu _{e}~$ can be written as%
\begin{equation}
{\cal L}^{\text{LFV}}_{\text{eff}}(\mu\text{-decay})=-\frac{G_{Y}}{\sqrt{2}}\left[ \bar{e}\gamma ^{\mu }(1+\gamma
^{5})\mu \right] \left[ \bar{\nu}_{e}\gamma _{\mu }(1-\gamma _{5})\nu _{\mu }%
\right] \label{36}
\end{equation}
and similarly for the $\tau $-decays, replace $\mu \rightarrow \tau $, $\nu _{\mu
}\rightarrow \nu _{\tau }$, and $\nu _{e}\rightarrow \nu _{\mu }$, in the above expression.

Using the permutation $D\left( 123\right) \rightarrow D\left(
231\right)$:
\begin{equation}
(ii) \hspace{1cm}D\left(
231\right)=\left( 
\begin{matrix}
\nu _{e} & \mu ^{+} \\ 
e^{-} & -N_{\mu }^{c}%
\end{matrix}%
\right) _{L}, ~\left( 
\begin{matrix}
\nu _{\mu } & \tau ^{+} \\ 
\mu ^{-} & -N_{\tau }^{c}%
\end{matrix}%
\right) _{L},~\left( 
\begin{matrix}
\nu _{\tau } & e^{+} \\ 
\tau ^{-} & -N_{e}^{c}%
\end{matrix}%
\right) _{L}
\end{equation}%
For the assignment $(ii)$, the lepton flavor violating decays are
\begin{align}
\mu ^{-} &\rightarrow \bar{\nu}_{e}+e^{-}+\nu _{\tau },		&\Delta L_{\mu
}&=-1, \hspace{1cm}\Delta L_{\tau }=1, \hspace{1cm}\nu _{\mu }\rightarrow \nu _{\tau } \\
\tau ^{-} &\rightarrow \bar{\nu}_{\mu }+\mu ^{-}+\nu _{e}, &\Delta L_{\tau
}&=-1,\hspace{1cm}\Delta L_{e}=1,\hspace{1cm}\nu _{\tau }\rightarrow ~\nu _{e} \\
\tau ^{-} &\rightarrow \bar{\nu}_{\mu }+e^{-}+\nu _{\tau },	&\Delta L_{\mu
}&=-1,\hspace{1cm}\Delta L_{e}=1,\hspace{1cm}\nu _{e}\rightarrow \nu _{\mu }
\end{align}%
These decays also stimulate the neutrino oscillatons $\nu _{\mu }\rightarrow \nu
_{\tau }$, $\nu _{\tau }\rightarrow ~\nu _{e}$, $\nu _{e}\rightarrow \nu
_{\mu }$ in $\mu $ and $\tau $ decays.
In this assignment $(ii)$, the charged LFV effective Lagrangian for the decay $\mu ^{-}\rightarrow \bar{\nu}%
_{e}+e^{-}+\nu _{\tau }$ is%
\begin{equation}
{\cal L}^{\text{LFV}}_{\text{eff}}(\mu\text{-decay})=-\frac{G_{Y}}{\sqrt{2}}[\bar{e}\gamma ^{\mu }(1+\gamma ^{5})\mu ][%
\bar{\nu}_{\tau }\gamma _{\mu }(1-\gamma _{5})\nu _{e}].  \label{effct5}
\end{equation}

The Feynman amplitude for the $\mu $ decay in the Standard Model is
given by [c.f. Eq. (\ref{eff-lag-1})] \cite{1}  
\begin{equation}
\big|M^{\text{SM}}_{\mu\text{-decay}}\big|^{2}=\sum_{\text{spin}}\big|F^{\text{SM}}_{\mu\text{-decay}}\big|^{2}\sim \frac{G_{F}^{2}}{2}~4~p_{2}\cdot k_{1}~p_{1}\cdot k_{2},
\label{amp1}
\end{equation}%
and for the lepton flavor violating $\mu $ decay is given by [c.f. Eq. (\ref{36})]
\begin{equation}
\big|M^{\text{LFV}}_{\mu\text{-decay}}\big|^{2}=\sum_{\text{spin}}\big|F^{\text{LFV}}_{\mu\text{-decay}}\big|^{2}\sim \frac{G_{Y}^{2}}{2}~4~p_{1}\cdot k_{1}~p_{2}\cdot k_{2},
\label{amp2}
\end{equation}%
where $p_{1},~p_{2},~k_{1}$ and $k_{2}$ are the 4-momenta of $\mu,~e,~\nu_{\mu
}$ and $\nu _{e}$.
From Eq. (\ref{amp1}), the decay width $d\Gamma $, for $\mu ^{-}\rightarrow
\nu _{\mu }+e^{-}+\bar{\nu}_{e}$ is given by [1]
\begin{equation}
d\Gamma^{\text{SM}}_{\mu\text{-decay}} =\frac{G_{F}^{2}}{12\pi ^{3}}m_{\mu
}p_{e}dE_{e}[3WE_e-2E_{e}^{2}-m_{e}^{2}], \hspace{1cm}\text{where}\hspace{1cm} 
W\equiv\frac{m_{\mu }^{2}+m_{e}^{2}}{2m_{\mu }} \label{drate}
\end{equation}%
After integration we get,
\begin{equation}
\Gamma^{\text{SM}}_{\mu\text{-decay}} =\tau _{\mu }^{-1}=\frac{G_{F}^{2}}{192\pi ^{3}}m_{\mu }^{5}[1-\frac{%
8m_{e}^{2}}{m_{\mu }^{2}}]. \label{trate}
\end{equation}%
From Eq. (\ref{amp2}), we get exactly the same expressions for the LFV case as those given in
Eq. (\ref{drate}) and Eq. (\ref{trate}) with $G_{F}^{2}\rightarrow G_{Y}^{2}$. For $\tau $ decays, replace $m_{\mu }\rightarrow m_{\tau }$, $%
m_{e}\rightarrow m_{\mu }$for $\tau \rightarrow \mu $ and for $\tau
\rightarrow e,$ $m_{\mu }\rightarrow m_{\tau }$. Hence, for the assignment
(i)
\begin{equation*}
{\cal R}^{\text{LFV}}_{\mu\text{-decay}}\equiv \frac{\Gamma^{\text{LFV}}_{\mu\text{-decay}} (\mu ^{-}\rightarrow \bar{\nu}_{\mu }+e^{-}+\nu _{e})}{\Gamma^{\text{SM}}_{\mu\text{-decay}}
(\mu ^{-}\rightarrow \nu _{\mu }+e^{-}+\bar{\nu}_{e})}=\frac{G_{Y}^{2}}{%
G_{F}^{2}}=\frac{g_{Y}^{4}}{g^{4}}\left( \frac{m_{W_{L}}}{m_{Y}}\right) ^{4} =\left( \frac{\sin ^{2}\theta _{W}}{1-2\sin ^{2}\theta _{W}}\right)^2 \left( 
\frac{m_{W_L}}{m_{Y}^{\prime }}\right) ^{4},
\end{equation*}%
Moreover, for the $\tau$-decay we get the same ratio, i.e. ${\cal R}^{\text{LFV}}_{\tau\text{-decay}}={\cal R}^{\text{LFV}}_{\mu\text{-decay}}$.
Similar expressions for the decays $\mu ^{-}\rightarrow \bar{\nu}%
_{e}+e^{-}+\nu _{\tau }$, $\tau ^{-}\rightarrow \bar{\nu}_{\mu }+\mu
^{-}+\nu _{e}$ and $\tau ^{-}\rightarrow \bar{\nu}_{\mu }+\mu ^{-}+\nu _{e}$
for the assignment (ii). The probability to observe lepton flavor violating $%
\mu $ and $\tau $ decays must be less than $10^{-6}$, since the SM decay
rate for $\mu $ decay is in agreement with the experimental value up to six
places of decimals. For example, using $\sin^{2}\theta _{W}=0.23$, $%
m_{W_{L}}\approx 80.38$ GeV, the probability to observe lepton flavor
violating decay is $2.9\times 10^{-8}$ for $m_{Y}=50m_{W_{L}}\approx 4$ TeV
and $1.5\times 10^{-9}$ for $m_{Y}=100m_{W_{L}}\approx 8$ TeV and for $%
m_{Y}=65$ TeV the probability is $5.0\times 10^{-13}.$

The energy spectrum given in Eq. (\ref{drate}) is modified by ${\cal L}^{\text{LFV}}_{\text{eff}}$ for
flavor violating decays. The Feynman amplitude for $\mu $ decay into
electron is given by 
\begin{equation}
\begin{split}
F^{\text{SM}}_{\mu\text{-decay}}& =\frac{G_{F}}{\sqrt{2}}\big[\bar{u}(p_{2})\gamma ^{\mu }(1-\gamma
^{5})u(p_{1})\big]\big[\bar{u}(k_{1})\gamma _{\mu }(1-\gamma _{5})v (k_{2})\big] \\
F^{\text{LFV}}_{\mu\text{-decay}}&= -\frac{G_{Y}}{\sqrt{2}}\big[\bar{u}(p_{2})\gamma ^{\mu }(1+\gamma
^{5})u(p_{1})\big]\big[\bar{u}(k_{2})\gamma _{\mu }(1-\gamma _{5})v (k_{1})\big],
\end{split}%
\end{equation}%
which leads to
\begin{equation}
|M_{\mu\text{-decay}}|^{2}\sim G_{F}^{2}~4~p_{2}\cdot k_{1}~p_{1}\cdot k_{2}-2G_{F}G_{Y}(-4m_{\mu
}m_{e}~k_{1}\cdot k_{2})  \label{amp-new}
\end{equation}%
From Eq. (\ref{amp-new}), the modified energy spectrum is given by 
\begin{equation}
d\Gamma =\frac{G_{F}^{2}}{12\pi ^{3}}m_{\mu
}p_{e}dE_{e}\big[3WE_{e}-2E_{e}^{2}-m_{e}^{2}-12(\frac{\sin ^{2}\theta _{W}}{%
1-\sin ^{2}\theta _{W}})\frac{m_{W_L}^{2}}{m_{Y}^{2}}m_{e}\times
(W-E_{e})\big].
\end{equation}

\section{Conclusions}
The electroweak unification group $SU(2)_{L}\times
SU(2)_{R}\times SU(2)_Y$ is broken to $SU(2)_{L}\times SU(2)_{R}\times
U(1)_{Y_{1}}$ by a scalar multiple $\Sigma $ that belongs to the triplet
representation of $SU(2)_Y$ and singlet of $SU(2)_{L}\times SU(2)_{R},$
the vector boson $Y^{\pm }$ acquired mass $m_{Y^{\pm }}^{2}=\frac{1}{2}\frac{%
\tan ^{2}\theta _{W}}{1-\tan ^{2}\theta _{W}}g^{2}v_\Sigma^{2} $,
where $v_\Sigma^{2}\gg v_{R}^{2}\gg v_{L}^{2}$. The vector bosons $Y^{\pm }$ are
decoupled, with one heavy Higgs scalar $H_{\Sigma }^{0}$ and the residual
group $SU(2)_{L}\times SU(2)_{R}\times U(1)_{Y_{1}}$ where $Y_{1}$ is the
hypercharge has all the features of the left-right symmetric electroweak
unification group$SU(2)_{L}\times SU(2)_{R}\times U(1)_{Y_{1}}$ with
hypercharge $Y_{1}=B-L$. $Y_{1}=B=\pm \frac{1}{3}$ for the quark multiplets
and $Y_{1}=-L=\mp 1$ for the lepton multiplets. Addition of quark multiplet
does not change any other feature of the group. The probability to observe
charged lepton flavor violating decays is $\leq 10^{-9}$.
\section*{Acknowledgment}
The author would like to thank Dr. Aqeel Ahmed for the helpful discussions.


\end{document}